\documentclass[12pt]{article} 
\usepackage{amsmath}
\usepackage{graphicx}
\usepackage{enumerate}
\usepackage{natbib}
\usepackage{multirow}
\usepackage{longtable}
\usepackage{booktabs}
\usepackage{amssymb}
\usepackage{bm}
\usepackage{mathrsfs}
\usepackage{amsthm}
\usepackage{amsfonts}
\usepackage{subfigure}
\usepackage{floatrow}
\usepackage{epsfig,amssymb,latexsym,verbatim}
\usepackage{epstopdf}
\usepackage{graphics}
\usepackage[english]{babel}
\usepackage[figuresright]{rotating}
\usepackage[dvipsnames]{xcolor}
\usepackage{color}
\usepackage{hyperref}
\usepackage[ruled,vlined,linesnumbered]{algorithm2e}
\usepackage{appendix}
\usepackage{adjustbox}
\usepackage{tabularx}
\usepackage{array}
\newcolumntype{P}[1]{>{\raggedright\arraybackslash\hspace{0pt}\baselineskip=9.2pt\relax}p{#1}}
\SetKwInput{KwInput}{Input}
\SetKwInput{KwOutput}{Output}
\SetAlgoSkip{smallskip}
\DontPrintSemicolon
\SetAlFnt{\small}
\SetAlCapFnt{\small}
\SetAlCapNameFnt{\small}
\SetKwComment{tcp*}{\hfill\(\triangleright\)\ }{}

\allowdisplaybreaks
\newtheorem{theorem}{Theorem}

\newtheorem{remark}{Remark}
\newtheorem{prop}{Proposition}

\numberwithin{equation}{section}
\numberwithin{lemma}{section}
\numberwithin{theorem}{section}
\numberwithin{prop}{section}
\numberwithin{corollary}{section}

\def\P{\mathbb{P}}
\def\E{\mathbb{E}}

\def\Cov{\mbox{Cov}}
\def\o{{\scriptstyle{\mathcal{O}}}}
\def\O{\mathcal{O}}

\def\diam{\text{diam}}

\def\bSig\mathbf{\Sigma}

\newcommand{\calM}{\mathcal{M}}

\newcommand{\blind}{0}

\begin{document}

\def\spacingset#1{\renewcommand{\baselinestretch}%
{#1}\small\normalsize}
\spacingset{1}

\date{}

\ifnum\blind=0\relax
{
  \title{\bf Multicollinearity-agnostic feature screening for
non-Euclidean responses: a factor adjusted approach}

  \renewcommand{\thefootnote}{\fnsymbol{footnote}}
  \author{Moshu Xu\footnotemark[1] \footnotemark[4] $\,$ Leheng Cai\footnotemark[1] \footnotemark[4] $\,$ Yanmei Shi\footnotemark[2] $\,$ Xu Guo\footnotemark[2] $\,$ and$\,$  Qirui Hu\footnotemark[3] \footnotemark[5]  \hspace{.2cm}}
  
  \renewcommand{\thefootnote}{\fnsymbol{footnote}}

  \footnotetext[1]{Department of Statistics and Data Science, Tsinghua University}
  \footnotetext[2]{School of Statistics, Beijing Normal University}
   \footnotetext[3]{School of Statistics and Data Science, Shanghai University of Finance and Economics}
   \footnotetext[4]{The first two authors contributed equally to this work. }
		\footnotetext[5]{Corresponding author: huqirui@mail.shufe.edu.cn}
    \maketitle
}
\else
{
  \bigskip
  \bigskip
  \bigskip
  \begin{center}
    {\bf Multicollinearity-agnostic feature screening for
non-Euclidean responses: a factor adjusted approach}
  \end{center}
  \medskip
}
\fi

\bigskip

\textbf{Abstract:} In high-dimensional settings, multicollinearity is a pervasive issue that can substantially impair the performance of feature screening methods based on marginal Fr\'echet regression.
Feature screening for non-Euclidean responses becomes unreliable when ultrahigh-dimensional predictors suffer from multicollinearity, because feature-specific signals may be masked by shared latent factors. To mitigate this effect, we propose a Factor adjusted Fr\'echet sure independence screening procedure. The method first recovers latent common factors from the predictors and then evaluates each feature by the incremental Fr\'echet coefficient of determination contributed by its idiosyncratic component beyond the common factors. Under regularity conditions, we establish uniform approximation rates for the feasible screening utilities and prove the sure screening and sure ranking properties. Extensive numerical experiments provide compelling empirical support for the validity and effectiveness of our approach, particularly in scenarios with highly correlated covariates. We further illustrate the practical performance of our method through two representative non-Euclidean datasets: the ADNI dataset and the mortality dataset, both with distribution-valued responses.


\newpage
\spacingset{1.9}

\section{Introduction}
\label{Section: Introduction}

\begin{figure}[ht]
    \centering
    \includegraphics[width=0.9\textwidth]{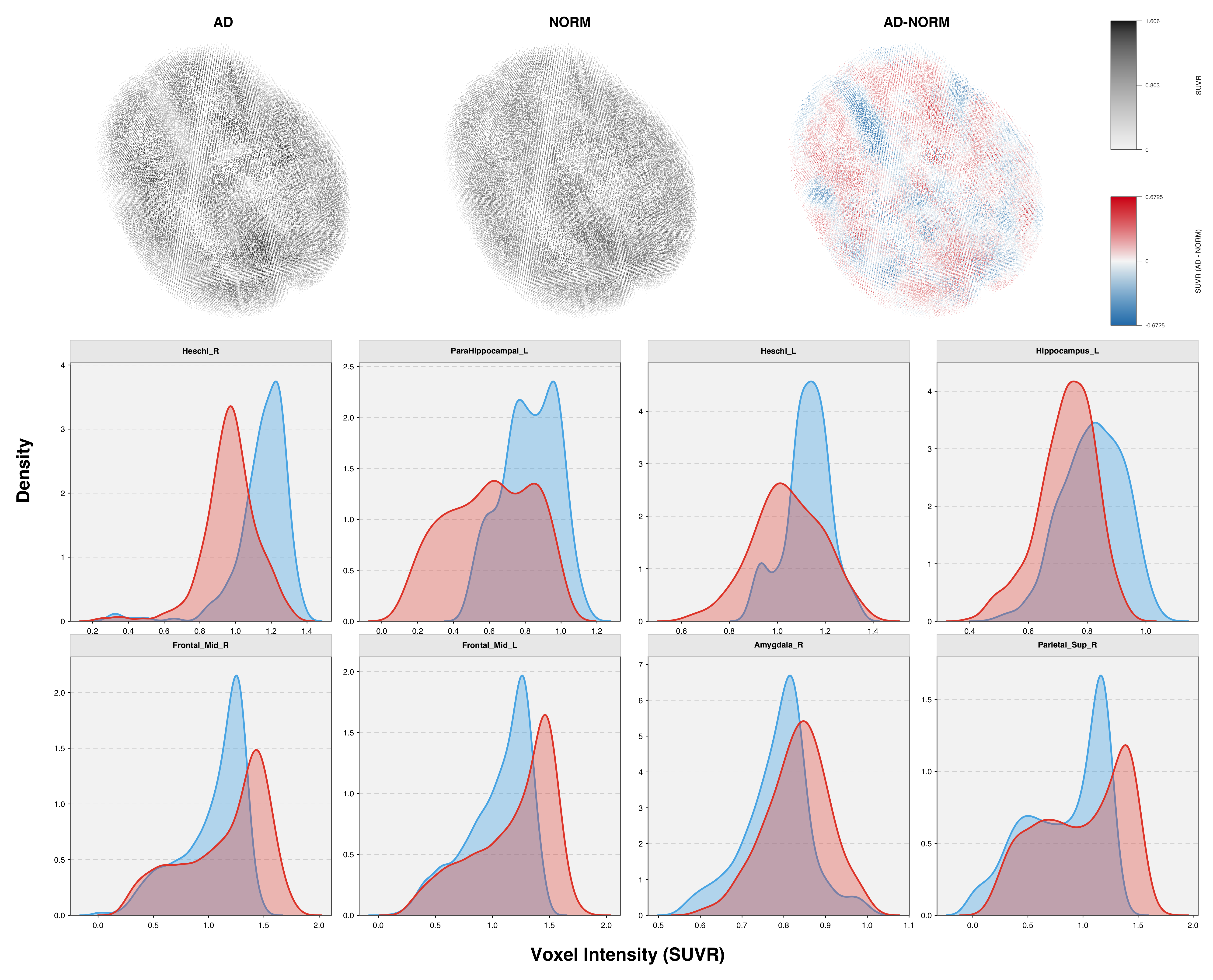}
    \caption{Top row: PET brain voxel maps for an AD subject, a cognitively normal 
    (NORM) subject, and their voxel-wise difference (AD–NORM), displayed after downsampling for visual clarity. Bottom two rows: voxel-intensity density functions for the same AD subject (red) and NORM subject (blue) in eight representative brain regions.}
    \label{Figure: Motivation 3 row}
\end{figure}

Alzheimer’s disease (AD) is a progressive neurodegenerative disorder whose pathological changes may begin years before clinical diagnosis. A central goal in modern imaging-genetics is therefore to identify genetic variants associated with early neuroimaging alterations \citep{Stein2010imaging, Shen2010imaging}. The Alzheimer’s Disease Neuroimaging Initiative (ADNI) provides longitudinal positron emission tomography (PET) measurements, together with genome-wide SNP data and clinical covariates, which substantially enriches relevant research; see \cite{Mueller2005ways}, \cite{Petersen2010clinical} and \cite{Weiner2013review} for details. The top row of Figure \ref{Figure: Motivation 3 row} gives a direct visual illustration of ADNI PET data. It displays downsampled PET voxel data for one AD subject and one cognitively normal subject, together with their difference map. The contrast suggests a scalar summary, such as the regional mean or a low-dimensional score, is potentially inadequate. To preserve more information, we adopt the approach advocated by \cite{nyul1999standardizing}, \cite{Goldsmith2012} and \cite{Zhongwei2025feature}, treating the responses as the probability density function of voxel intensities within specific brain regions.
The bottom two rows of Figure \ref{Figure: Motivation 3 row} depict the voxel intensity density functions of the two subjects in the top row across eight brain regions. The differences involve not only location shifts but also changes in spread and shape. These plots motivate treating each regional imaging outcome as a distribution-valued object, which naturally places the response in a non-Euclidean metric space such as the Wasserstein space. Meanwhile, the predictors include tens of thousands of SNPs plus clinical variables, placing us in an ultrahigh-dimensional regime where feature screening is needed.

To address non-Euclidean responses, Object-Oriented Data Analysis has gained significant traction \citep{marron2014overview}. A fundamental tool in this domain is Fr\'echet regression, which generalizes the classical conditional mean to metric spaces \citep{frechet1948elements, MullerFrechet}. When the predictor dimension is ultrahigh, however, regression must usually be preceded by feature screening. Sure independence screening and its model-free extensions provide scalable tools for Euclidean responses \citep{FanSIS, zhu2011model, li2012feature}. More recently, \cite{Zhongwei2025feature} proposed a Global Fr\'echet SIS procedure, utilizing the marginal Fr\'echet coefficient of determination to rank the effect of each SNP on ADNI brain density curves. Related recent developments include variable screening in Fr\'echet regression for diffusion tensor imaging \citep{yan2025variable} and conditional screening procedures for non-Euclidean responses \citep{gao2025identification}.

Despite these advances, existing metric-space screening methods typically assume predictors are weakly correlated. However, in genomic studies, linkage disequilibrium and population structure induce
pervasive dependence \citep{khvorykh2025determinant} that is often well-approximated by a low-dimensional latent factor structure \citep{de2022multivariate}. As a result, ordinary marginal screening tends to select redundant proxies and may miss genuinely informative variables whose marginal signals are partially masked by the common factor structure. A toy example illustrates the issue. Suppose that in the setting of linear regression, the high-dimensional predictor vector follows a factor model, with each covariate decomposed into a common factor part and an idiosyncratic part. Imagine that the response depends on the latent common factors and also on the idiosyncratic components of several truly active covariates. It may happen that an inactive covariate, simply because it loads strongly on the same common factor direction, exhibits a larger marginal association with the response than a truly active covariate whose factor loading partially offsets that direction; see the  detailed mathematical descriptions in Section S.1 of the Supplementary Material.

For Euclidean responses, a substantial literature addresses screening under factor-type dependence. \cite{wang2012factor} proposed a factor-profiled screening method. \cite{kneip2011factor} and \cite{fan2020factor} developed factor-adjusted
 regularized
  selection methods. \cite{Fan2024latent} proposed a factor augmented regression  model (FARM), which removes the common factor effects from covariates before fitting and variable selection. However, extending these ideas to metric-space responses is not immediate.  Fr\'echet regression is defined via minimization of expected squared distances rather than linear projections, and standard Euclidean arguments based on residual additivity and covariance algebra do not directly transfer.

To bridge this gap, we propose a novel Factor adjusted Fr\'echet sure independence screening procedure, referred to as Factor adjusted SIS. This method extends the robustness of factor adjusted screening to the general metric space setting. We assume that the predictor vector follows a high-dimensional factor model \(X_i = \mathbf Bf_i + u_i\), where \(f_i\) collects the common latent factors, $\mathbf B$ is the loading matrix and \(u_i\) is the idiosyncratic component. The factor structure is first estimated by principal component analysis \citep{Stock2002, Bai2003}. For each feature \(k\), we then compare two global Fr\'echet regressions: a baseline model using the common factors \(f_i\) alone, and an augmented model using \((f_i^\top,u_{ik})^\top\). The importance of feature \(k\) is quantified by the additional Fr\'echet coefficient of determination contributed by \(u_{ik}\) beyond the common factors. In this way, the proposed utility measures whether the \(k\)th predictor contains information about the object-valued response that is not already explained by the latent dependence structure shared across predictors. 

Our main contributions are threefold. First, we propose a Factor adjusted SIS procedure for responses taking values in a general metric space. The method isolates the unique contribution of each covariate beyond the common factor structure and avoids Euclidean residualization of the response and therefore respects the intrinsic geometry of the response space. 
Second, we develop a modular theoretical analysis for the proposed procedure. We establish an exponential concentration bound for the sample Fr\'{e}chet coefficient of determination, quantify the impact of factor recovery on the screening utility, and show uniform consistency of the feasible incremental utilities, which in turn yields the sure screening property and, under an additional signal separation condition, the sure ranking property. 
Third, through extensive simulations and two real data analyses, we demonstrate that the proposed method is particularly effective under strong multicollinearity, significantly improving screening accuracy while delivering more interpretable and less redundant discoveries.

The remainder of the paper is arranged as follows. In Section~\ref{Section: Review}, we briefly review the global Fr\'echet regression and derive the statistical properties of the associated coefficient of determination $R_{\oplus}^2$. In Section~\ref{Section: Factor adjusted Frechet SIS}, we introduce the screening procedure and establish its theoretical guarantees. Section~\ref{Section: Simulation} and section~\ref{Section: Real data analysis} demonstrate the effectiveness of the proposed method via simulation studies and real data analysis, respectively. Concluding remarks are given in Section~\ref{Section: Concluding remark}.

\section{Review of the global Fr\'echet regression and the $R_{\oplus}^2$}
\label{Section: Review}

This section reviews the fixed-dimensional Fr\'echet regression method in \cite{MullerFrechet}, which underlies our screening procedure. The predictor in this section is denoted by $Z\in\mathbb{R}^q$, where $q$ is fixed. In Section~\ref{Section: Factor adjusted Frechet SIS}, the generic predictor $Z$ will be instantiated as the common factor vector $f\in\mathbb{R}^K$ or as the augmented vector $(f^\top,u_{k})^\top \in \mathbb{R}^{K+1}$, where $u_k$ is the $k$-th idiosyncratic component. Thus, Section~\ref{Section: Review} develops the generic module, while Section~\ref{Section: Factor adjusted Frechet SIS} specifies how this module is used for ultrahigh-dimensional screening.

Consider a random pair $(Y,Z)$, where $Y$ takes values in a metric space $(\Omega,d)$ and $Z\in\mathbb{R}^q$ has support $\mathcal{Z}\subset\mathbb{R}^q$, mean $\mu_Z$, and nonsingular covariance matrix $\Sigma_Z$. The unconditional Fr\'echet mean and variance of $Y$ are defined as $\omega_{\oplus} = \arg\min_{\omega \in \Omega} \E[d^2(Y, \omega)]$ and $V_{\oplus} = \E[d^2(Y, \omega_{\oplus})]$, respectively. To model the regression relationship between $Y$ and $Z$, \cite{MullerFrechet} define the global Fr\'echet regression, which generalizes the classical multiple linear regression to the metric space setting. For $z,t\in\mathcal Z$, let $s(z,t) = 1 + (z-\mu_Z)^\top \Sigma_Z^{-1}(t-\mu_Z)$. The regression function $m_{\oplus}(t)$ is then defined as the minimizer of a weighted Fr\'echet objective function
\begin{align}
\label{Equation: Generic m and M}
    m_{\oplus}(t) = \arg\min_{\omega \in \Omega}M(\omega,t), \qquad M(\omega,t) = \E[s(Z, t)d^2(Y, \omega)].
\end{align}
We assume throughout that each global Fr\'echet regression used to define the population screening utilities is correctly specified, in the sense that $m_{\oplus}(t)$ is the unique conditional Fr\'echet mean of $Y$ given $Z=t$ for $P_Z$-almost every $t$.
To quantify the explanatory power of the regression, we use the coefficient of determination associated with the predictor $Z$: 
\begin{align}
\label{Equation: Generic R2}
    R_\oplus^2 = 1 - \frac{\E[d^2(Y, m_{\oplus}(Z))]}{V_\oplus}.
\end{align}
Under the preceding setting, $0\leq R_\oplus^2\leq1$. Moreover, $R_\oplus^2$ cannot decrease when the predictor vector is enlarged, since conditioning on more information cannot increase the minimum conditional squared-distance risk. 

Given i.i.d. observations $\{(Y_i,Z_i)\}_{i=1}^n$, let $\bar Z = n^{-1} \sum_{i=1}^n Z_i$, $\widehat\Sigma_Z = n^{-1} \sum_{i=1}^n (Z_i-\bar Z)(Z_i-\bar Z)^\top$, and define the empirical weight $\widehat s(z,t) = 1 + (z-\bar Z)^\top \widehat\Sigma_Z^{-1}(t-\bar Z)$. The sample Fr\'echet regression function is obtained by minimizing the sample objective
\begin{align}
\label{Equation: Generic sample m and M}
    \widehat{m}_{\oplus}(t) = \arg\min_{\omega \in \Omega} \widehat{M}(\omega,t), \qquad \widehat{M}(\omega,t) = \frac{1}{n} \sum_{i=1}^n \widehat{s}(Z_i, t)d^2(Y_i, \omega).
\end{align}
The sample analogue of (\ref{Equation: Generic R2}) is
\begin{align}
\label{Equation: Generic sample R2}
    \widehat R_\oplus^2 = 1 - \frac{\sum_{i=1}^n d^2 \left(Y_i,\widehat m_{\oplus}(Z_i)\right)}{\sum_{i=1}^n d^2(Y_i,\widehat\omega_\oplus)},
\end{align}
where $\widehat\omega_\oplus = \arg\min_{\omega\in\Omega} n^{-1} \sum_{i=1}^n d^2(Y_i,\omega)$.

To derive a valid screening utility index based on $R_{\oplus}^2$, it must possess certain properties. We impose the following regularity assumptions and derive an exponential tail bound for $R_{\oplus}^2$.
\begin{itemize}
    \item[(A1)] The metric space $(\Omega,d)$ is bounded, i.e. $\diam(\Omega) <\infty$. The Fr\'echet mean $\omega_{\oplus}$ exists and is unique. There exists a constant $v>0$ such that Fr\'echet variance $V_{\oplus}\geq v>0$.
    \item[(A2)] $Z-\mu_Z$ is sub-Gaussian in $\mathbb R^q$, namely there exists a constant $K_Z<\infty$, such that $\sup_{\|u\|_E=1} \|u^\top(Z-\mu_Z) \|_{\psi_2}\le K_Z$,  where $\|\cdot\|_E$ denotes the Euclidean norm. The covariance matrix $\Sigma_Z$ is nonsingular and $\lambda_{\min}(\Sigma_Z) \geq \lambda_0 >0$. 
    \item[(A3)] For each $t\in\mathbb{R}^q$, $m_{\oplus}(t)$ exists and is unique. For all $n$, $\widehat{m}_{\oplus}(t)$ exists and is unique almost surely. Additionally, for every $\epsilon >0$,
    \begin{align*}
        \inf_{t\in \mathbb{R}^q}\inf_{d(\omega, m_{\oplus}(t)) >\epsilon} \{ M(\omega, t) - M(m_{\oplus}(t),t) \}>0.
    \end{align*}
    \item[(A4)] There exist constants $\eta_1>0$, $C_1>0$, and $\beta_1>1$ such that, uniformly over $t\in\mathbb R^{q}$,
    \begin{align*}
        M(\omega,t) - M(m_{\oplus}(t),t) \geq C_1 d(\omega, m_{\oplus}(t))^{\beta_1}
    \end{align*}
    provided $d(\omega, m_{\oplus}(t)) <\eta_1$. 
    \item[(A5)] Let $B_c(r)=\{\omega\in\Omega:d(\omega,c)\le r\}$ and let $N(r,A,d)$ denote the covering number of a subset $A\subset \Omega$ under metric $d$. There exist constants $J_c<\infty$, $a_\Omega\ge0$, and $b_\Omega>0$ such that, for every $0<\epsilon\leq 1$,  
    \begin{align*}
        \int_0^1 \sqrt{\log N(u\epsilon,\Omega,d)}du \leq a_{\Omega} + b_{\Omega}\sqrt{\log(1/\epsilon)},\qquad \sup_{c\in \Omega}\int_0^1 \sqrt{\log N(u\epsilon, B_c(2\epsilon),d)}du \leq J_c .
    \end{align*} 
\end{itemize}
Assumptions (A1) and (A3) are standard well-posedness and identification conditions for
global Fr\'echet regression. They exclude pathological cases in which the population or sample
Fr\'echet regression mean is not uniquely defined or cannot be separated from competing
objects away from the minimizer. These conditions are standard in works on object data analysis, such as \cite{bhattacharya2003large}, \cite{MullerFrechet} and \cite{bhattacharjee2023single}. The simulation results suggest that the proposed procedure remains effective in unbounded response spaces such as $\mathbb{R}$ and the family of Gaussian distributions under the Wasserstein metric. Assumption (A2) is used to control the sample mean, the sample covariance matrix, and the resulting regression weights. Assumption (A4) is the key device that converts uniform stochastic fluctuations of the objective function into metric
deviations of the regression estimator. Similar local convexity requirements can be found in \cite{thomas2013geodesic} and \cite{tucker2023variable}. Assumption (A5) provides the
uniform empirical process control required for the weighted Fr\'echet objective. Analogous conditions are used in \cite{MullerFrechet} and \cite{dubey2020functional}. We use an example in Euclidean space to illustrate how the above assumptions are satisfied.
\begin{prop}
\label{Proposition: Euclidean example}
    Let $\Omega$ be a bounded, closed and convex subset of $\mathbb{R}^l$ equipped with the Euclidean metric $d_E(Y_1,Y_2)=\|Y_1-Y_2\|_2$. Assume $Y\in\Omega$ almost surely. Then, assumptions (A3)-(A5) hold with $\beta_1 = 2$, $C_1 = 1$, $\eta_1 = \infty$, $a_\Omega=\sqrt{l\log(1+2\diam(\Omega))}+\sqrt{\pi l}/2$, $b_{\Omega} = \sqrt{l}$ and $J_c = \sqrt{l\log 5}+\sqrt{\pi l}/2$.
\end{prop}

\begin{remark}
\label{Remark: Assumption A1H}
The boundedness of $\Omega$ in (A1) is a convenient envelope condition rather than an intrinsic restriction of the proposed framework. A sufficient replacement is (A1H): $(\Omega,d)$ is isometric to a nonempty closed convex subset $C$ of a separable Hilbert space $\mathcal H$, and, for some fixed $\omega\in\Omega$ and $K_Y<\infty$, $\E\exp\left\{d^2(Y,\omega) / K_Y^2 \right\}\leq 2$. Under (A1H), with (A2)--(A5) unchanged, the conclusions in this paper remain valid. Closedness and convexity ensure well-defined Hilbert projections, while the tail condition replaces boundedness in the concentration arguments. The tail condition is automatic when $\Omega$ is bounded. This condition covers Euclidean and $L^2$-valued responses, positive semidefinite matrices with the Frobenius distance, $\mathcal{W}_2(\mathbb R)$ via quantiles, and the univariate Gaussian--Wasserstein family with a closed scale set. For this family, the scale set may be either $\sigma\geq 0$ or $\sigma \geq \sigma_{\min} > 0$.

\end{remark}

The next result gives an exponential concentration bound for $\widehat R_\oplus^2$.
\begin{theorem}
\label{Theorem: R2 tail bound}
    Under Assumptions (A1)-(A5), there exist constants $c_R, C_R>0$ and $\eta_\star\in(0,\eta_1\wedge 1]$, depending only on the regularity constants in Assumptions (A1)-(A5), such that, for any $0<\epsilon< \eta_\star$ and for all sufficiently large $n$, we have
    \begin{align*}
        \mathbb{P}\left( | \widehat{R}_{\oplus}^2 - R_{\oplus}^2 | \geq \epsilon \right) \leq c_R \exp{\left(-C_Rn\min\left\{
\epsilon^2,\,
\frac{\epsilon^{2(\beta_1-1)}}{1+\log(1/\epsilon)}
\right\}\right)} .
    \end{align*}
\end{theorem}
The role of Theorem~\ref{Theorem: R2 tail bound} is purely modular. It establishes an exponential concentration bound for $\widehat R_\oplus^2$, which can be regarded as a building block in following analysis. In Section~\ref{Section: Factor adjusted Frechet SIS}, although the ambient covariate dimension is ultrahigh, each feature-specific regression involves only a low-dimensional set of predictors with size $q=K$ or $q=K+1$. Hence, Theorem~\ref{Theorem: R2 tail bound} remains applicable to  each oracle feature-wise regression.

\section{Methodologies}
\label{Section: Factor adjusted Frechet SIS}
\subsection{The Factor adjusted SIS}
For an ultrahigh-dimensional covariate vector $X_i=(X_{i1},\dots,X_{ip})^\top \in \mathbb{R}^p$, assume that
\begin{align*}
    X_i = \mathbf Bf_i + u_i, \qquad i=1,\dots,n,
\end{align*}
where $\mathbf B\in\mathbb{R}^{p\times K}$ is the loading matrix, $f_i\in\mathbb{R}^K$ is the vector of common factors, and $u_i=(u_{i1},\dots,u_{ip})^\top\in\mathbb{R}^p$ is the idiosyncratic component. In matrix notation, $\mathbf{X} = \mathbf{F} \mathbf{B}^{\top} + \mathbf{U}$.  Throughout, $K$ is fixed, whereas $p$ is allowed to diverge with $n$.

To remedy the spurious correlation caused by common factors in the Euclidean-response setting, \cite{Fan2024latent} adopted a two-step estimation method: first regress $\mathbf Y$ on $\mathbf{F}$ to obtain residuals $\widetilde{\mathbf Y} = \mathbf Y - P_{\mathbf{F}}\mathbf Y$, where $P_{\mathbf{F}}$ is the projection matrix. Then fit a model for $\widetilde{\mathbf Y}$ and the idiosyncratic components $\mathbf{U}$. Since $\mathbf Y = (\mathbf{F}\mathbf{B}^{\top} + \mathbf{U})^{\top}\bm{\beta} + \epsilon$ in the presence of factor model structure, the coefficient of $\mathbf{U}$ should be identical to that of $\mathbf{X}$. However, this residual-based strategy encounters a fundamental bottleneck in the non-Euclidean framework. We cannot perform vector subtraction $\mathbf Y - P_{\mathbf{F}}\mathbf Y$ in spaces lacking a linear structure. If we define a residual as $d(\mathbf Y, P_{\mathbf{F}}\mathbf Y)$, the result is a non-negative scalar. Using scalars to replace non-Euclidean objects destroys the geometric structure of $\Omega$ and results in a severe loss of information. 

To address the non-Euclidean nature of $Y$ while adjusting for the latent factors in $\mathbf{X}$, we propose assessing the importance of the $k$-th feature via model comparison. The baseline model regresses $Y$ only on common factors, while the $k$-th augmented model regresses $Y$ on common factors and the $k$-th idiosyncratic component. The explanatory power of the $k$-th feature for $Y$ is represented by the difference in $R_{\oplus}^2$ between the two models.

We now use a unified notation to represent these two models. The key point is that we \emph{do not} introduce a new regression model in this section. Instead, we use the generic regression function and $R_{\oplus}^2$ from Section \ref{Section: Review} with chosen fixed-dimensional predictors. Specifically, let
\begin{align*}
    Z_{i0}=f_i\in\mathbb{R}^K, \qquad Z_{ik}=\left(f_i^\top,u_{ik}\right)^\top \in \mathbb{R}^{K+1}, \qquad k=1,\dots,p.
\end{align*}
Write $Z_k$ for a generic random vector having the same distribution as $Z_{ik}$. Applying the construction of Section~\ref{Section: Review} to $Z_k$ yields the population regression function
\begin{align}
\label{Equation: Population mk and Mk}
    m_{\oplus,k}(t) = \arg\min_{\omega \in \Omega} M_k(\omega,t), \qquad M_k(\omega,t) = \E[s_k(Z_k, t)d^2(Y, \omega)],
\end{align}
for $t\in \mathrm{supp}(Z_k)$ and $k=0,\dots,p$. Note that $s_k(z_k,t) = 1 + (z_k -\mu_k)^\top \Sigma_k^{-1}(t-\mu_k)$, where $z_k\in \mathrm{supp}(Z_k)$, $\mu_k = \E Z_k$ and $\Sigma_k = \Cov(Z_k)$. The corresponding population utility is
\begin{align}
\label{Equation: Population Rk}
    R_{\oplus,k}^2 = 1 - \frac{\E[d^2(Y, m_{\oplus,k}(Z_k))]}{V_\oplus}, \qquad k=0,\dots, p.
\end{align}
Here $R_{\oplus,0}^2$ is the explanatory power of the factor-only model, while $R_{\oplus,k}^2$ for $k\ge 1$ is the explanatory power of the model based on $Z_k$. The additional contribution of the $k$-th covariate after adjusting for the common factors is measured by the incremental utility
\begin{align*}
    \Delta_k = R_{\oplus,k}^2 - R_{\oplus,0}^2, \qquad k=1,\dots,p.
\end{align*}
A large value of $\Delta_k$ indicates that the idiosyncratic component of $X_k$ contributes explanatory power beyond the latent factors. Consequently, we use $\Delta_k$ as the screening index.

In practice, neither $f_i$ nor $u_{ik}$ is observed. Hence, we use the
following PCA-based estimator and its oracle version for \eqref{Equation: Population mk and Mk} and \eqref{Equation: Population Rk}. Let $\widehat K$ be a consistent estimator of $K$. We write the formulas below as if $\widehat K=K$. Let $\widehat f_i$ and $\widehat u_{ik}$ be the
PCA-based estimators of $f_i$ and $u_{ik}$, respectively. Define the feasible predictors by
\begin{align*}
    \widehat Z_{i0}=\widehat f_i, \qquad \widehat{Z}_{ik} =(\widehat f_i^\top, \widehat u_{ik})^\top, \qquad k=1,\dots,p.
\end{align*}
We now apply the same sample regression function and $R_{\oplus}^2$ from (\ref{Equation: Generic sample m and M}) and (\ref{Equation: Generic sample R2}) at two different inputs: the oracle predictor $Z_{ik}$ and the feasible predictor $\widehat{Z}_{ik}$. This gives that, for $k=0,\dots, p$,
\begin{align*}
    \widetilde{m}_{\oplus,k}(t) = \arg\min_{\omega \in \Omega}\widetilde{M}_k(\omega, t), \qquad \widetilde{M}_k(\omega,t) = \frac{1}{n}\sum_{i=1}^n \widetilde{s}_k(Z_{ik}, t)d^2(Y_i, \omega),\\
    \widehat{m}_{\oplus,k}(t) = \arg\min_{\omega \in \Omega}\widehat{M}_k(\omega, t), \qquad \widehat{M}_k(\omega,t) = \frac{1}{n}\sum_{i=1}^n \widehat{s}_k(\widehat{Z}_{ik}, t)d^2(Y_i, \omega),
\end{align*}
and 
\begin{align*}
    \widetilde{R}_{\oplus,k}^2 = 1 - \frac{\sum_{j=1}^n d^2\left(Y_j,\widetilde m_{\oplus,k}(Z_{jk})\right)}{\sum_{j=1}^n d^2(Y_j,\widehat\omega_\oplus)}, \qquad \widehat{R}_{\oplus,k}^2 = 1 - \frac{\sum_{j=1}^n d^2\left(Y_j,\widehat m_{\oplus,k}(\widehat{Z}_{jk})\right)}{\sum_{j=1}^n d^2(Y_j,\widehat\omega_\oplus)}.
\end{align*}
Note that $\widetilde{R}_{\oplus,k}^2$ and $\widehat{R}_{\oplus,k}^2$ are calculated by fitting at the sample points. The tilde notation indicates the oracle statistic computed from the unobserved predictor $\{Z_{ik}\}_{i=1}^n$, whereas the hat notation indicates the feasible statistic obtained by $\left\{\widehat Z_{ik}\right\}_{i=1}^n$. In other words, Section~\ref{Section: Review} always provides the same building block. The only difference is whether its input predictor is oracle or estimated. Correspondingly, define the oracle and feasible sample incremental utilities by
\begin{align*}
    \widetilde{\Delta}_k = \widetilde{R}_{\oplus,k}^2 - \widetilde{R}_{\oplus,0}^2, \qquad \widehat\Delta_k=\widehat R_{\oplus,k}^2-\widehat R_{\oplus,0}^2, \qquad k=1,\dots, p.
\end{align*}

We can now formally introduce the screening rule. Define the set of true active predictors $\mathcal{M}$ and inactive predictors $\mathcal{I}$ as
\begin{align*}
    \mathcal{M} = \{ 1\leq k\leq p: \Delta_k >0 \},\qquad
    \mathcal{I} = \{ 1,\dots, \}\setminus\mathcal{M}.
\end{align*}
For any $X_k\in\mathcal{M}$, $X_k$ contains predictive information about the conditional Fr\'echet mean of $Y$ that is not captured by the common factors. For a given threshold $\tau_{n,p}$, we retain
\begin{align*}
    \widehat{\mathcal M} = \{1\le k\le p: \widehat\Delta_k \geq \tau_{n,p}\}.
\end{align*}
The theoretical order of $\tau_{n,p}$ is discussed in detail in Theorem~\ref{Theorem: Combined with factor model}.
This screening rule is aligned with the residual-based estimation strategy in \cite{Fan2024latent} and our model comparison idea, that is to retain variables whose idiosyncratic components improve the factor-only Fr\'echet regression. Algorithm~\ref{Algorithm: Factor adjusted SIS} illustrates the concrete procedure of the proposed Factor adjusted SIS. Factor estimation can be implemented by a truncated SVD of the $n\times p$ predictor matrix. Computing the leading $K_{\max}+1$ components for the eigenvalue-ratio criterion requires approximately $O(npK_{\max})$ operations. The factor-only fit is computed once, after which the screening cost grows linearly with $p$. The feature-wise fits can be divided into predictor blocks and run in parallel.

\begin{algorithm}[ht]
\caption{Factor adjusted Fr\'echet Sure Independence Screening}
\label{Algorithm: Factor adjusted SIS}
\small
\renewcommand{\baselinestretch}{1.0}\selectfont
\setlength{\parskip}{0pt}
\textbf{Input:} Data $\{(Y_i,X_i)\}_{i=1}^n$ with $X_i=(X_{i1},\ldots,X_{ip})^\top$, metric space $(\Omega,d)$, and threshold $\tau_{n,p}$.\\
\textbf{Output:} The estimated active set $\widehat{\calM}$.

\textbf{Step 1: Factor recovery.} 
Estimate the number of factors $\widehat K$ and obtain PCA-based estimators $\widehat f_i$ and $\widehat u_i=(\widehat u_{i1},\ldots,\widehat u_{ip})^\top$, $i=1,\ldots,n$.

\textbf{Step 2: Initialization.} 
Compute the sample Fr\'echet mean
$\widehat\omega_\oplus=\arg\min_{\omega\in\Omega} n^{-1}\sum_{i=1}^n d^2(Y_i,\omega)$
and set $\mathrm{SST}=\sum_{i=1}^n d^2(Y_i,\widehat\omega_\oplus)$.

\textbf{Step 3: Utility estimation.}
Construct the factor-only predictors $\widehat Z_{i0}=\widehat f_i$, $i=1,\ldots,n$, and compute the corresponding sample global Fr\'echet regression fit $\widehat m_{\oplus,0}$ and utility
$\widehat R_{\oplus,0}^2
=
1-\sum_{i=1}^n d^2\!\left(Y_i,\widehat m_{\oplus,0}(\widehat Z_{i0})\right)/\mathrm{SST}$.

\For{$k=1,\ldots,p$}{
Construct the augmented predictors $\widehat Z_{ik}=\left(\widehat f_i^\top,\widehat u_{ik} \right)^\top$, $i=1,\ldots,n$.

Compute the sample global Fr\'echet regression fit $\widehat m_{\oplus,k}$ based on $\left\{ \left(Y_i,\widehat Z_{ik} \right)\right\}_{i=1}^n$.

Compute
$\widehat R_{\oplus,k}^2
=
1-\sum_{i=1}^n d^2\!\left(Y_i,\widehat m_{\oplus,k}(\widehat Z_{ik})\right)/\mathrm{SST}$,
and set
$\widehat\Delta_k=\widehat R_{\oplus,k}^2-\widehat R_{\oplus,0}^2$.
}

\textbf{Step 4: Screening.}
Set
$\widehat{\calM}
=
\{k:\widehat\Delta_k\ge \tau_{n,p}\}$.

\Return{$\widehat{\calM}$}.
\end{algorithm}

\subsection{Theoretical results}
We now state the conditions governing factor recovery and signal strength.
\begin{itemize}
    \item[(B1)] $\widehat K$ is a consistent estimator of $K$.
    \item[(B2)]  The factors and idiosyncratic components are centered and uniformly sub-Gaussian. Moreover, $\E(f_if_i^\top)=I_K$, $\E(f_iu_i^\top)=0$.
    \item[(B3)] The eigenvalues of $\mathbf B^\top \mathbf B/p$ are bounded away from zero and infinity, and there exists $\Upsilon>0$ such that $\|\mathbf B\|_{\max}\le \Upsilon$, $\E|u_i^\top u_i-\operatorname{tr}(\Sigma_u)|^4\le \Upsilon p^2$, where $\Sigma_u=\operatorname{Cov}(u_i)$. There exists $\kappa\in(0,1)$ such that $\kappa \leq \lambda_{\min}(\Sigma_u)\leq \lambda_{\max}(\Sigma_u)\le \kappa^{-1}$, $\|\Sigma_u\|_1\le \kappa^{-1}$ and $\min_{1\le k,\ell\le p}\operatorname{Var}(u_{ik}u_{i\ell})\ge \kappa$.
    \item[(B4)] Let $\Sigma = \Cov(X)$ and $\widehat{\Sigma} = n^{-1}\sum_{i=1}^n (X_i - \bar{X})(X_i - \bar{X})^{\top}$. Let $\Lambda = \mathrm{diag}(\lambda_1, \dots, \lambda_K)$ and $\Gamma = (\zeta_1, \dots, \zeta_K)$ denote the matrix of leading $K$ eigenvalues of $\Sigma$ and the corresponding matrix of orthonormal eigenvectors, respectively. Let $\widehat{\Lambda}$ and $\widehat{\Gamma}$ be their estimators based on $\widehat{\Sigma}$. Then $\left\| \widehat{\Sigma} - \Sigma \right\|_{\max} = \O_p \left(\sqrt{(\log p)/n} \right)$, $\left\| (\widehat{\Lambda} - \Lambda)\widehat{\Lambda}^{-1} \right\|_{\max} = \O_p \left(\sqrt{(\log p)/n} \right)$ and $\left\| \widehat{\Gamma} - \Gamma \right\|_{\max} = \O_p \left(\sqrt{(\log p)/(np)} \right)$.  
\end{itemize}
Assumptions (B1)-(B4) are standard in high-dimensional factor models \citep{Bai2003, fan2013large, li2018embracing, Fan2024latent}.
Assumption (B1) guarantees consistent recovery of the factor dimension. We use the eigenvalue-ratio method numerically, while the theory treats $K$ as known, conditional on the consistency of $\widehat K$. Assumptions (B2)--(B3) impose sub-Gaussianity, orthogonality, bounded pervasive loadings, and a regular idiosyncratic covariance, ensuring model identifiability and PCA consistency. Assumption (B4) provides the uniform convergence of $\widehat f_i$ and $\widehat u_i$ needed to compare the Fr\'echet objectives based on $Z_{ik}$ and $\widehat Z_{ik}$. Since $s(z,t)$ is invariant under nonsingular linear transformations of $z$, the factor rotation is absorbed into the factors and loadings and omitted from the notation.

\begin{itemize}
    \item [(C1)] $\log p = \o(\sqrt{n})$ and $\log(n) = \o(p)$.
    \item [(C2)] The minimum signal strength $\min_{k\in \mathcal{M}} \Delta_k > 2\tau_{n,p}$.
\end{itemize}
Assumption (C1) restricts the dimensional growth. Assumption (C2) is a minimum signal condition commonly employed in feature screening procedures, as seen, for example, in \cite{pan2019generic, guo2022stable, he2024large}. It requires that the utility of active predictors dominate the screening threshold, and that the threshold in turn dominate the stochastic error of the screening statistic. Theorem~\ref{Theorem: Combined with factor model} specifies the required order of $\tau_{n,p}$.

\begin{theorem}
\label{Theorem: Combined with factor model}
    Suppose that Assumptions (A1)-(A5) hold for each regression based on $Z_{k}$, $k=0,\dots,p$, and that the constants appearing in these assumptions can be chosen uniformly in $k$. Assume further that Assumptions (B1)-(B4) and (C1) hold. Then, conditional on event 
    \begin{gather*}
        E_n = \{ \inf_{0\leq k\leq p} \inf_{1\leq j\leq n} \inf_{d(\omega, \widetilde{m}_{\oplus,k}(Z_{jk})) \geq \eta_0 } \widetilde{M}_{k}(\omega, Z_{jk}) - \widetilde{M}_{k}(\widetilde{m}_{\oplus, k}(Z_{jk}),Z_{jk}) > \delta_0\\
        \inf_{0\leq k\leq p} \inf_{1\leq j\leq n} \inf_{d(\omega, \widetilde{m}_{\oplus,k}(Z_{jk})) < \eta_0} \widetilde{M}_{k}(\omega, Z_{jk}) - \widetilde{M}_{k}(\widetilde{m}_{\oplus, k}(Z_{jk}),Z_{jk}) \geq c_0 d(\omega, \widetilde{m}_{\oplus, k}(Z_{jk}))^{\beta_1} \}
    \end{gather*}
    with constants $\eta_0, c_0, \delta_0>0$, we have
    \begin{align*}
        \max_{1\leq k\leq p} \left| \widehat{\Delta}_k - \Delta_k \right| = \O_p(r_{n,p}),
    \end{align*}
    where
    \begin{align*}
        r_{n,p} := \left( \sqrt{\frac{\log (np)}{p}} + \sqrt{\frac{(\log p)(\log (np))}{n}} \right)^{\frac{1}{\beta_1}} .
    \end{align*}
    Further, if the minimum signal strength condition (C2) holds with $r_{n,p} = \o(\tau_{n,p})$, we can obtain the sure screening property
    \begin{align*}
        \mathbb{P}\left( \mathcal{M} \subseteq \widehat{\mathcal{M}} \right) \to 1,\qquad  n\to\infty.
    \end{align*}
\end{theorem}
In conclusion, Theorem~\ref{Theorem: Combined with factor model} guarantees that with probability approaching one, all true active predictors are retained in the reduced model $\widehat{\mathcal{M}}$. 

\begin{remark}
\label{Remark: En}
$E_n$ is a uniform separation and curvature condition for the oracle criteria $\widetilde M_k(\cdot,Z_{jk})$, which  converts the uniform closeness of the oracle criteria $\widetilde{M}_k$ and feasible criteria $\widehat{M}_k$ into closeness of their minimizers. In the Euclidean setting of Proposition \ref{Proposition: Euclidean example}, this event holds automatically with $\beta_1=2$ and $c_0=1$. Under (A1H) in Remark~\ref{Remark: Assumption A1H}, $E_n$ holds with with $\beta_1=2$ and $c_0=1$ whenever the sample covariance matrices used in the oracle weights are nonsingular. This covers one-dimensional Wasserstein responses and the closed univariate Gaussian--Wasserstein family.
For Riemannian responses, $E_n$ holds when the response space is restricted to a compact strongly geodesically convex region, the squared-distance Hessians are uniformly positive there, and the total negative weight is sufficiently small. This covers small spherical balls and bounded geodesically convex subsets of Hadamard manifolds.
\end{remark}

In practice, feature screening is commonly implemented by either choosing those predictors whose $\widehat{\Delta}_k$ exceeds a threshold, like in Algorithm~\ref{Algorithm: Factor adjusted SIS}, or ranking the predictors by descending $\widehat{\Delta}_k$ and retaining the top $d_n$ predictors. $d_n$ is a user-specified integer that grows with $n$ but remains much smaller than $p$. While the threshold based approach is theoretically justified by Theorem~\ref{Theorem: Combined with factor model}, the following result validates the top-$d_n$ selection.

\begin{itemize}
    \item[(C3)] $\min_{k\in \mathcal{M}} \Delta_k - \max_{k\in \mathcal{I}} \Delta_k = \Delta_{\mathrm{rank}} >0$, $r_{n,p} = \o(\Delta_{\mathrm{rank}})$. 
\end{itemize}
This assumption formally posits a separation in the utility index between active and inactive predictors, a key condition for ranking based screening methods and widely adopted in high-dimensional feature screening \citep{guo2022stable, li2024feature, Zhongwei2025feature}. Note that $\Delta_{\mathrm{rank}}$ can tend to $0$ as $n\to \infty$.

\begin{theorem}
\label{Theorem: Sure ranking property}
Under the settings in Theorem~\ref{Theorem: Combined with factor model} and Assumption (C3), we have the following sure ranking property
    \begin{align*}
        \P\left( \min_{k\in \mathcal{M}} \widehat{\Delta}_k \geq \max_{k\in \mathcal{I}} \widehat{\Delta}_k \right) \to 1.
    \end{align*}
\end{theorem}
Theorem~\ref{Theorem: Sure ranking property} establishes that Factor adjusted SIS possesses the sure ranking property \citep{zhu2011model}. It ranks all active predictors above inactive ones with probability tending to one.

\section{Numerical results}
\label{Section: Simulation}

In this section, we evaluate the finite-sample performance of the proposed Factor adjusted SIS method through simulation studies. We compare our method with three competitors: Global Fr\'echet SIS \citep{Zhongwei2025feature}, Distance Correlation SIS (DC-SIS; \citealp{li2012feature}) and Ball Correlation SIS (BCor-SIS; \citealp{pan2019generic}).
We consider three cases corresponding to Euclidean data, distribution-valued data, and spherical data. For all settings, we fix the number of factors $K=2$, the active covariate set to be $\{1, 3, 5 \}$ and the size of the active set $|\mathcal{M}|=3$. The sample size is set to be $n \in \{100, 200\}$ and the dimension of covariates is $p \in \{1000, 2000\}$. For each $(n,p)$ combination, we generate $200$ Monte Carlo replications.
We assess the screening performance of all methods using the following criteria: $P_k$, the proportion of simulations in which a specific active variable $X_k$ is selected, $P_{\mathrm{all}}$, the proportion of simulations in which all active predictors are successfully recruited, and the distribution of the minimum model size required to include all active features, reported via its mean and $5\%, 25\%, 50\%, 75\%,$ and $95\%$ quantiles. Additional simulations are reported in Section S.2 of the Supplementary Material, where we conduct stress tests on spherical responses. The scenarios cover factor misspecification, incorrect numbers of factors, and heavy-tailed and correlated idiosyncratic errors.

The predictors $X_i \in \mathbb{R}^p$ are generated from a latent factor model $X_i = \mathbf Bf_i + u_i$, where the common factors $f_i \sim N_K(0, I_K)$ and the idiosyncratic errors $u_i \sim N_p(0, I_p)$. The entries of the factor loading matrix $\mathbf B$ are drawn independently from $U(-1,1)$. This structure ensures that $X$ has a pervasive dependence structure.
We estimate $\mathbf B$, $\{f_i\}_{i=1}^n$, $\{u_i\}_{i=1}^n$ and the number of common factors $K$ according to the factor model literature \cite{Bai2003}, \cite{Stock2002} and \cite{Fan2024latent}. Let $\widehat{K}$ denote a consistent estimator of the number of factors $K$, which is estimated by the eigenvalue ratio method. The estimator for the factor matrix, $\widehat{\mathbf{F}}$, is constructed such that its columns are $\sqrt{n}$ times the eigenvectors corresponding to the $\widehat{K}$ largest eigenvalues of the matrix $\mathbf{X}\mathbf{X}^\top$ and $\widehat{\mathbf{B}} = n^{-1}\mathbf{X}^\top \widehat{\mathbf{F}}$. The resulting estimator for the idiosyncratic component matrix $\mathbf{U}$ is given by the projection residuals $\widehat{\mathbf{U}} = \mathbf{X} - \widehat{\mathbf{F}}\widehat{\mathbf{B}}^\top = \left(\mathbf{I}_n - \widehat{\mathbf{P}}\right)\mathbf{X}$, where $\widehat{\mathbf{P}} = n^{-1}\widehat{\mathbf{F}}\widehat{\mathbf{F}}^\top$ is the projection matrix onto the space spanned by the estimated factors.

\subsection{Case 1: $Y\in \mathbb{R}$}

\begin{figure}[ht]
    \centering  \includegraphics[width=0.9\textwidth]{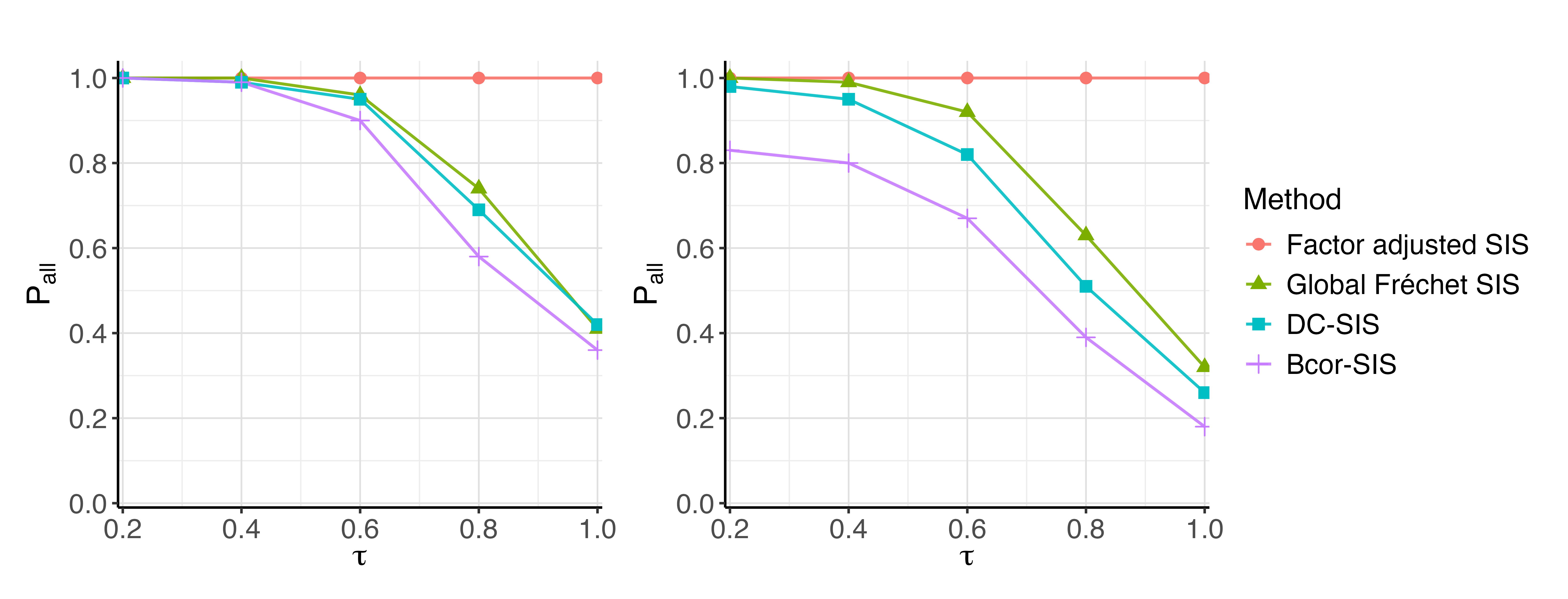} 
    \caption{Under the settings $\sigma_{\epsilon}=0.5$ (left) and $\sigma_{\epsilon}=1$ (right), the probability that all active variables are screened by each of the four methods under different values of $\tau$.} 
    \label{Figure: Simulation case1 comparison} 
\end{figure}

In the first scenario, we consider Euclidean responses. The responses are generated from $Y_i = X_i^{\top} \beta + \varepsilon_i$, where $\varepsilon_i \sim N(0, \sigma_{\epsilon}^2)$. When generating $X$, we additionally multiply the factor loading matrix $\mathbf B$ by a scale parameter $\tau$, i.e. $X_i = \tau \mathbf Bf_i + u_i$. This allows us to investigate the performance of our method under different levels of factor structure strength. We set $\sigma_{\epsilon} = 0.5$ and $\tau = 1$ in the setting of Table~\ref{Table: Simulation case 1}, and $\sigma_{\epsilon} = 0.5, 1$, $\tau = 0.2, 0.4, 0.6, 0.8, 1$ in the setting of Figure~\ref{Figure: Simulation case1 comparison}. We set the active entries of $\beta$ to $0.5$, i.e. $\beta_1 = \beta_{3} = \beta_{5} = 0.5$, while all its remaining entries are zero. The coefficients and noise level were chosen to maintain a moderate signal-to-noise ratio, ensuring that the true signals are recoverable only if the confounding effects of $f$ are properly adjusted.

\subsection{Case 2: $Y\in \mathcal{W}_2$}

In this scenario, the response objects are univariate probability distributions, treated as elements in the Wasserstein space $\mathcal{W}_2$. We generate $Y_i$ as Gaussian distributions parameterized by their quantile functions $Q_i(\tau_j) = \mu_i + \sigma_i \Phi^{-1}(\tau_j)$ on $(\tau_1, \dots, \tau_{m-1}) = (1/m, 2/ m, \dots, 1-1/m)$ with $m=25$.  
The location $\mu_i$ and scale $\sigma_i$ are determined by
\begin{align*}
    &\mu_i|X_i \sim N(\mu_0 + \beta_{\mu}(X_{i3}+X_{i5}), \nu_1),\\
    &\sigma_i|X_i \sim \mathrm{Gamma}\left(m_{\sigma,i}^2/\nu_2,\, \nu_2/m_{\sigma,i}\right), \qquad m_{\sigma,i}=\sigma_0\exp\left(\beta_{\sigma}X_{i1}/\sigma_0\right).
\end{align*}
This log-link guarantees $m_{\sigma,i}>0$ for all $X_{i1}$. In the implementation we take $\mu_0 = 0$, $\sigma_0 = 3$, $\beta_{\mu}=\beta_{\sigma}=1$ and $\nu_1 = \nu_2 = 0.5^2$.

\subsection{Case 3: $Y \in \mathbb{S}^2$}

This scenario tests the method on a nonlinear Riemannian manifold. The response $Y_i^{\mathrm{true}} \in \mathbb{S}^2$ is generated via a spiral curve wrapped around the sphere, that is 
\begin{align*}
    Y_i^{\mathrm{true}} = \left( \sqrt{1-\bar{X}_i^2}\cos(\pi \bar{X}_i),  \sqrt{1-\bar{X}_i^2} \sin(\pi \bar{X}_i), \bar{X}_i \right)^\top,
\end{align*}
where $\bar{X}_i$ is obtained by rescaling $(X_{i1} + X_{i3} + X_{i5})/3$ to the range $(0, 1)$. To generate the observed data $Y_i$, we take 2D Gaussian noise $V_i$ on the tangent space of $Y_i^{\mathrm{true}}$, and map it back to the sphere using the exponential map
\begin{gather*}
    Y_i = \cos(\|V_i\|_E)Y_i^{\mathrm{true}} + \sin(\|V_i\|_E)\frac{V_i}{\|V_i\|_E}.
\end{gather*}  
We set the variance of $V_i$ to be $0.5^2$. This construction yields a tightly concentrated distribution around the spiral curve while preserving the spherical geometry.

\subsection{Results}
Tables~\ref{Table: Simulation case 1}--\ref{Table: Simulation case 3} report the results for the three cases. Across all cases, Factor adjusted SIS attains the largest probabilities of recovering the active predictors and the smallest minimum model size. The advantage is particularly clear when the predictor dimension is large and when the signal structure is more challenging, as in Case 3. 
Figure~\ref{Figure: Simulation case1 comparison} presents the probability that all active variables are selected by the four methods under different values of $\sigma_{\epsilon}$ and factor structure strength $\tau$. The performance of all methods improves as the factor strength $\tau$ decreases. The identification rate for all methods is $1$ when the factor structure is very weak,  which validates the effectiveness of our Factor adjusted SIS in simple scenarios.
Tables S.1--S.6 in the Supplementary Material show that Factor adjusted SIS remains generally robust under moderate model misspecification. The largest losses occur under heavy-tailed or correlated errors, but its performance improves as the sample size increases.

\begin{table}[ht]
\caption{Selection probabilities and summary statistics of the minimum model size for Case 1 over $200$ replications.}
\label{Table: Simulation case 1} 
\centering
\renewcommand{\arraystretch}{0.65}
\resizebox{\textwidth}{!}{%
\begin{tabular}{@{}cc lrrrrrrrrrr@{}}
\toprule
$p$ & $n$ & Method & $P_{1}$ & $P_{3}$ & $P_{5}$ & $P_{\text{all}}$ &
$5\%$ & $25\%$ & $50\%$ & $75\%$ & $95\%$ & Mean \\
\midrule
\multirow{8}{*}{1000}
& \multirow{4}{*}{100}
& \textbf{Factor adjusted SIS} & \textbf{1.00} & \textbf{0.99} & \textbf{1.00} & \textbf{0.99} & \textbf{3.00} & \textbf{3.00} & \textbf{3.00} & \textbf{3.00} & \textbf{5.00} & \textbf{3.36} \\
& & Global Fr\'echet SIS          & 0.76 & 0.73 & 0.75 & 0.33 & 3.00 & 9.00 & 75.50 & 233.20 & 774.30 & 185.49 \\
& & DC-SIS                      & 0.75 & 0.73 & 0.76 & 0.33 & 3.00 & 12.00 & 84.00 & 281.00 & 749.00 & 188.72 \\
& & BCor-SIS                    & 0.71 & 0.66 & 0.68 & 0.22 & 3.95 & 25.00 & 130.50 & 375.25 & 720.10 & 223.31 \\
\cmidrule(l){2-13}
& \multirow{4}{*}{200}
& \textbf{Factor adjusted SIS} & \textbf{1.00} & \textbf{1.00} & \textbf{1.00} & \textbf{1.00} & \textbf{3.00} & \textbf{3.00} & \textbf{3.00} & \textbf{3.00} & \textbf{3.00} & \textbf{3.00} \\
& & Global Fr\'echet SIS          & 0.81 & 0.84 & 0.83 & 0.52 & 3.00 & 4.00 & 31.50 & 289.80 & 623.20 & 223.31 \\
& & DC-SIS                      & 0.81 & 0.82 & 0.80 & 0.49 & 3.00 & 3.75 & 42.50 & 293.50 & 655.60 & 168.56 \\
& & BCor-SIS                    & 0.80 & 0.78 & 0.80 & 0.44 & 3.00 & 7.00 & 77.00 & 306.00 & 699.00 & 190.17 \\
\midrule
\multirow{8}{*}{2000}
& \multirow{4}{*}{100}
& \textbf{Factor adjusted SIS} & \textbf{0.99} & \textbf{0.99} & \textbf{1.00} & \textbf{0.99} & \textbf{3.00} & \textbf{3.00} & \textbf{3.00} & \textbf{3.00} & \textbf{5.00} & \textbf{3.60} \\
& & Global Fr\'echet SIS          & 0.69 & 0.73 & 0.75 & 0.27 & 3.00 & 17.50 & 158.50 & 520.80 & 1445.00 & 372.80 \\
& & DC-SIS                      & 0.65 & 0.70 & 0.72 & 0.21 & 3.00 & 31.20 & 175.50 & 586.00 & 1424.40 & 391.30 \\
& & BCor-SIS                    & 0.58 & 0.63 & 0.67 & 0.14 & 5.95 & 57.25 & 260.50 & 771.25 & 1527.35 & 468.98 \\
\cmidrule(l){2-13}
& \multirow{4}{*}{200}
& \textbf{Factor adjusted SIS} & \textbf{1.00} & \textbf{1.00} & \textbf{1.00} & \textbf{1.00} & \textbf{3.00} & \textbf{3.00} & \textbf{3.00} & \textbf{3.00} & \textbf{3.00} & \textbf{3.00} \\
& & Global Fr\'echet SIS          & 0.80 & 0.76 & 0.82 & 0.41 & 3.00 & 4.00 & 116.00 & 552.00 & 1602.00 & 355.16 \\
& & DC-SIS                      & 0.79 & 0.76 & 0.83 & 0.42 & 3.00 & 4.75 & 111.50 & 521.50 & 1650.95 & 366.67 \\
& & BCor-SIS                    & 0.76 & 0.75 & 0.79 & 0.36 & 3.00 & 12.80 & 197.50 & 644.00 & 1565.90 & 408.17 \\
\bottomrule
\end{tabular}%
}
\end{table}

\begin{table}[ht]
\caption{Selection probabilities and summary statistics of the minimum model size for Case 2 over $200$ replications.}
\label{Table: Simulation case 2} 

\centering
\renewcommand{\arraystretch}{0.65}
\resizebox{\textwidth}{!}{%
\begin{tabular}{@{}cc lrrrrrrrrrr@{}}
\toprule
$p$ & $n$ & Method & $P_{1}$ & $P_{3}$ & $P_{5}$ & $P_{\text{all}}$ &
$5\%$ & $25\%$ & $50\%$ & $75\%$ & $95\%$ & Mean \\
\midrule
\multirow{8}{*}{1000}
 & \multirow{4}{*}{100}
 & \textbf{Factor adjusted SIS} & \textbf{1.00} & \textbf{1.00} & \textbf{1.00} & \textbf{1.00} & \textbf{3.00} & \textbf{3.00} & \textbf{3.00} & \textbf{3.00} & \textbf{3.00} & \textbf{3.04} \\
 &  & Global Fr\'echet SIS          & 0.86 & 0.85 & 0.86 & 0.60 & 3.00 & 3.00 & 12.00 & 56.20 & 279.10 & 55.17 \\
 &  & DC-SIS                      & 0.94 & 0.84 & 0.88 & 0.68 & 3.00 & 3.00 & 6.00 & 35.20 & 266.10 & 41.47 \\
 &  & BCor-SIS                    & 0.98 & 0.81 & 0.86 & 0.67 & 3.00 & 3.00 & 6.50 & 49.00 & 306.60 & 53.30 \\
\cmidrule(l){2-13}
 & \multirow{4}{*}{200}
 & \textbf{Factor adjusted SIS} & \textbf{1.00} & \textbf{1.00} & \textbf{1.00} & \textbf{1.00} & \textbf{3.00} & \textbf{3.00} & \textbf{3.00} & \textbf{3.00} & \textbf{3.00} & \textbf{3.00} \\
 &  & Global Fr\'echet SIS          & 0.91 & 0.95 & 0.90 & 0.77 & 3.00 & 3.00 & 3.00 & 29.80 & 231.00 & 43.23 \\
 &  & DC-SIS                      & 0.97 & 0.95 & 0.91 & 0.84 & 3.00 & 3.00 & 3.00 & 13.00 & 165.00 & 31.00 \\
 &  & BCor-SIS                    & 0.97 & 0.94 & 0.92 & 0.83 & 3.00 & 3.00 & 3.00 & 11.00 & 184.00 & 32.37 \\
\midrule
\multirow{8}{*}{2000}
 & \multirow{4}{*}{100}
 & \textbf{Factor adjusted SIS} & \textbf{1.00} & \textbf{1.00} & \textbf{1.00} & \textbf{1.00} & \textbf{3.00} & \textbf{3.00} & \textbf{3.00} & \textbf{3.00} & \textbf{4.00} & \textbf{3.15} \\
 &  & Global Fr\'echet SIS          & 0.84 & 0.86 & 0.87 & 0.61 & 3.00 & 3.00 & 7.00 & 91.00 & 479.00 & 88.72 \\
 &  & DC-SIS                      & 0.92 & 0.87 & 0.88 & 0.69 & 3.00 & 3.00 & 6.00 & 56.50 & 321.50 & 61.15 \\
 &  & BCor-SIS                    & 0.94 & 0.85 & 0.86 & 0.65 & 3.00 & 3.00 & 6.00 & 53.20 & 370.10 & 72.68 \\
\cmidrule(l){2-13}
 & \multirow{4}{*}{200}
 & \textbf{Factor adjusted SIS} & \textbf{1.00} & \textbf{1.00} & \textbf{1.00} & \textbf{1.00} & \textbf{3.00} & \textbf{3.00} & \textbf{3.00} & \textbf{3.00} & \textbf{3.00} & \textbf{3.00} \\
 &  & Global Fr\'echet SIS          & 0.90 & 0.94 & 0.94 & 0.78 & 3.00 & 3.00 & 3.00 & 30.20 & 457.50 & 65.64 \\
 &  & DC-SIS                      & 0.95 & 0.95 & 0.93 & 0.83 & 3.00 & 3.00 & 3.00 & 9.25 & 272.65 & 40.41 \\
 &  & BCor-SIS                    & 0.96 & 0.93 & 0.92 & 0.82 & 3.00 & 3.00 & 3.00 & 12.00 & 215.00 & 36.60 \\
\bottomrule
\end{tabular}%
}
\end{table}

\begin{table}[ht]
\caption{Selection probabilities and summary statistics of the minimum model size for Case 3 over $200$ replications.}
\label{Table: Simulation case 3} 

\centering
\renewcommand{\arraystretch}{0.65}
\resizebox{\textwidth}{!}{%
\begin{tabular}{@{}cc lrrrrrrrrrr@{}}
\toprule
$p$ & $n$ & Method & $P_{1}$ & $P_{3}$ & $P_{5}$ & $P_{\text{all}}$ &
$5\%$ & $25\%$ & $50\%$ & $75\%$ & $95\%$ & Mean \\
\midrule
\multirow{8}{*}{1000}
 & \multirow{4}{*}{100}
 & \textbf{Factor adjusted SIS} & \textbf{0.78} & \textbf{0.78} & \textbf{0.79} & \textbf{0.52} & \textbf{3.00} & \textbf{6.00} & \textbf{20.00} & \textbf{95.80} & \textbf{366.00} & \textbf{86.33} \\
 &  & Global Fr\'echet SIS          & 0.61 & 0.65 & 0.59 & 0.16 & 6.00 & 43.00 & 114.00 & 374.00 & 807.00 & 235.62 \\
 &  & DC-SIS                      & 0.58 & 0.61 & 0.58 & 0.12 & 8.00 & 51.50 & 142.50 & 397.20 & 798.30 & 250.94 \\
 &  & BCor-SIS                    & 0.47 & 0.47 & 0.47 & 0.04 & 26.00 & 101.00 & 209.00 & 456.00 & 802.00 & 302.69 \\
\cmidrule(l){2-13}
 & \multirow{4}{*}{200}
 & \textbf{Factor adjusted SIS} & \textbf{0.99} & \textbf{0.98} & \textbf{0.96} & \textbf{0.93} & \textbf{3.00} & \textbf{3.00} & \textbf{3.00} & \textbf{5.25} & \textbf{66.05} & \textbf{13.66} \\
 &  & Global Fr\'echet SIS          & 0.73 & 0.75 & 0.77 & 0.34 & 3.00 & 16.80 & 132.00 & 362.50 & 726.60 & 224.03 \\
 &  & DC-SIS                      & 0.73 & 0.74 & 0.76 & 0.33 & 3.00 & 19.00 & 146.00 & 398.00 & 730.00 & 238.25 \\
 &  & BCor-SIS                    & 0.68 & 0.69 & 0.68 & 0.19 & 6.00 & 50.50 & 196.50 & 481.50 & 883.60 & 295.08 \\
\midrule
\multirow{8}{*}{2000}
 & \multirow{4}{*}{100}
 & \textbf{Factor adjusted SIS} & \textbf{0.70} & \textbf{0.69} & \textbf{0.72} & \textbf{0.37} & \textbf{3.95} & \textbf{9.00} & \textbf{49.50} & \textbf{192.75} & \textbf{667.50} & \textbf{153.25} \\
 &  & Global Fr\'echet SIS          & 0.60 & 0.52 & 0.62 & 0.09 & 9.95 & 74.00 & 235.50 & 785.25 & 1611.50 & 471.59 \\
 &  & DC-SIS                      & 0.56 & 0.53 & 0.60 & 0.08 & 14.00 & 93.50 & 252.00 & 846.20 & 1712.50 & 511.21 \\
 &  & BCor-SIS                    & 0.41 & 0.38 & 0.44 & 0.04 & 24.00 & 140.00 & 426.00 & 1039.00 & 1720.00 & 619.73 \\
\cmidrule(l){2-13}
 & \multirow{4}{*}{200}
 & \textbf{Factor adjusted SIS} & \textbf{0.96} & \textbf{0.96} & \textbf{0.97} & \textbf{0.90} & \textbf{3.00} & \textbf{3.00} & \textbf{3.00} & \textbf{8.00} & \textbf{161.00} & \textbf{29.33} \\
 &  & Global Fr\'echet SIS          & 0.70 & 0.74 & 0.70 & 0.28 & 3.00 & 30.00 & 200.00 & 646.00 & 1537.00 & 440.69 \\
 &  & DC-SIS                      & 0.68 & 0.71 & 0.71 & 0.26 & 3.00 & 36.50 & 232.00 & 740.80 & 1723.2 & 471.66 \\
 &  & BCor-SIS                    & 0.62 & 0.60 & 0.65 & 0.12 & 14.90 & 91.20 & 340.50 & 863.00 & 1569.70 & 533.56 \\
\bottomrule
\end{tabular}%
}
\end{table}

\section{Real data analysis}
\label{Section: Real data analysis}
We illustrate the practical value of Factor adjusted SIS in two applications. The first is an imaging-genetics study from ADNI, where the predictors are hundreds of thousands of SNPs and strong dependence arises from linkage disequilibrium and population structure. The second is a cross-country mortality study, where many macro-level indicators are highly collinear because they reflect overlapping aspects of socioeconomic development. In both settings, the goal is to identify features that make distinct contributions beyond dominant latent common factors.

\subsection{Case 1: ADNI data}
The Alzheimer’s Disease Neuroimaging Initiative (ADNI) is a longitudinal study designed to identify neuroimaging, genetic, and clinical biomarkers of Alzheimer’s disease progression. After integrating PET imaging, genotype and clinical covariates, our working sample contains 235 subjects: 69 cognitively normal controls, 117 subjects with mild cognitive impairment, and 49 patients with Alzheimer’s disease, with imaging acquired at baseline, month 6, and month 12. The predictors consist of 582591 SNPs and five clinical covariates: age, sex, body weight, NPISCORE, and FAQSCORE. Each SNP is a single-nucleotide variation at a genomic locus, with values 0 (homozygous for major allele), 1 (heterozygous), and 2 (homozygous for minor allele).

The brain is partitioned into 42 regions using the Automated Anatomical Labeling (AAL) atlas. As motivated by Figure~\ref{Figure: Motivation 3 row}, we represent each region by the distribution of PET voxel intensities to preserve more information. Before screening, we retain SNPs with a call rate of at least $0.95$ and a minor allele count (MAC) of at least 10, following conventional genotype quality-control practice \citep{anderson2010data} and other single-variant analyses of Alzheimer’s disease data \citep{bis2020whole}. Age and sex are included as adjustment covariates in the baseline Fr\'echet regression, together with the estimated common factors. For each brain region and each visit, we apply Factor adjusted SIS and summarize the results by screening frequency across analyses. Our implementation used R 4.3.0, blocks of 4000 SNPs, and BLAS matrix operations. The single-process analysis of 547603 quality-controlled SNPs across 42 brain regions and three visits takes approximately 26 minutes on an 8-core Apple M2 laptop with 16 GB RAM.

Table~\ref{Table: Top 20 SNPs and frequency} presents the top 20 most frequently screened SNPs across regions and time points, along with their screening frequencies. 
The most frequently selected SNP, rs10204084, is located in an intron of gene DPP10. The same SNP was selected in another ADNI analysis of the time from mild cognitive impairment to Alzheimer's disease \citep{lee2023bayesian}. Abnormal DPP10 expression has also been observed in neurofibrillary tangles and plaque-associated dystrophic neurites in Alzheimer's disease brains \citep{chen2014dpp10}. Several other selected SNPs are located in genes that have been studied in relation to Alzheimer's disease. SNP rs2030490 is located in gene KIAA2012, which is associated with progression from mild cognitive impairment to Alzheimer's disease in both ADNI and NACC \citep{liu2022epigenetic}. SNP rs1542226 is located in gene NTM, and another variant in this gene has been linked to tau pathology \citep{chibnik2018susceptibility}. SNP rs4677602 is located in gene FOXP1. Studies in neuronal cells and mice found that this gene regulates a known Alzheimer's disease risk gene \citep{fu2025foxp1}. Therefore, our method may provide a novel tool and perspective for AD research.
However, the biological evidence above concerns different levels, including SNP, gene, and related mechanisms. Hence, we suggest viewing these SNPs as promising candidates rather than confirmed genetic associations with Alzheimer's disease. Further validation in larger and independent cohorts is needed.

\begin{table}[ht]
\centering
\caption{Rank and frequency of the top 20 SNPs across regions and time points.}
\label{Table: Top 20 SNPs and frequency}
\renewcommand{\arraystretch}{0.65}
\begin{tabular}{rlr @{\hskip 0.8cm} rlr}
\toprule
Rank & SNP & Frequency & Rank & SNP & Frequency \\
\midrule
1  & rs10204084 & 56 & 11 & rs4906122  & 34 \\
2  & rs9550900  & 50 & 12 & rs2157663  & 32 \\
3  & rs2419629  & 47 & 13 & rs4677602  & 32 \\
4  & rs6911282  & 40 & 14 & rs2030490  & 30 \\
5  & rs253721   & 39 & 15 & rs10786910 & 30 \\
6  & rs4802605  & 37 & 16 & rs8008197  & 28 \\
7  & rs7211577  & 36 & 17 & rs1275582  & 28 \\
8  & rs628844   & 36 & 18 & rs10860561 & 25 \\
9  & rs1542226  & 36 & 19 & rs2379054  & 24 \\
10 & rs669776   & 34 & 20 & rs4794117  & 22 \\
\bottomrule
\end{tabular}
\end{table}

\subsection{Case 2: Mortality data}
We next analyze female mortality distributions across countries. The response is the age-at-death distribution for females in a given country and year, constructed from the United Nations World Population Prospects (WPP) table “Deaths by single age – female,” which records deaths at each age from 0 to 100. The predictors are country-level indicators from the World Bank World Development Indicators (WDI). The data are available from \url{https://datatopics.worldbank.org/world-development- indicators} and \url{https://population.un.org/wpp}. After matching the two sources, the analysis includes 80 countries in 2021 and 388 indicators, spanning economy and national accounts, social and labor conditions, health, agriculture and rural development, governance and institutions, environment and energy, infrastructure, technology, and education. 

\begin{figure}[htbp]
    \centering
    \begin{minipage}{0.41\textwidth}
        \centering
        \includegraphics[width=\linewidth]{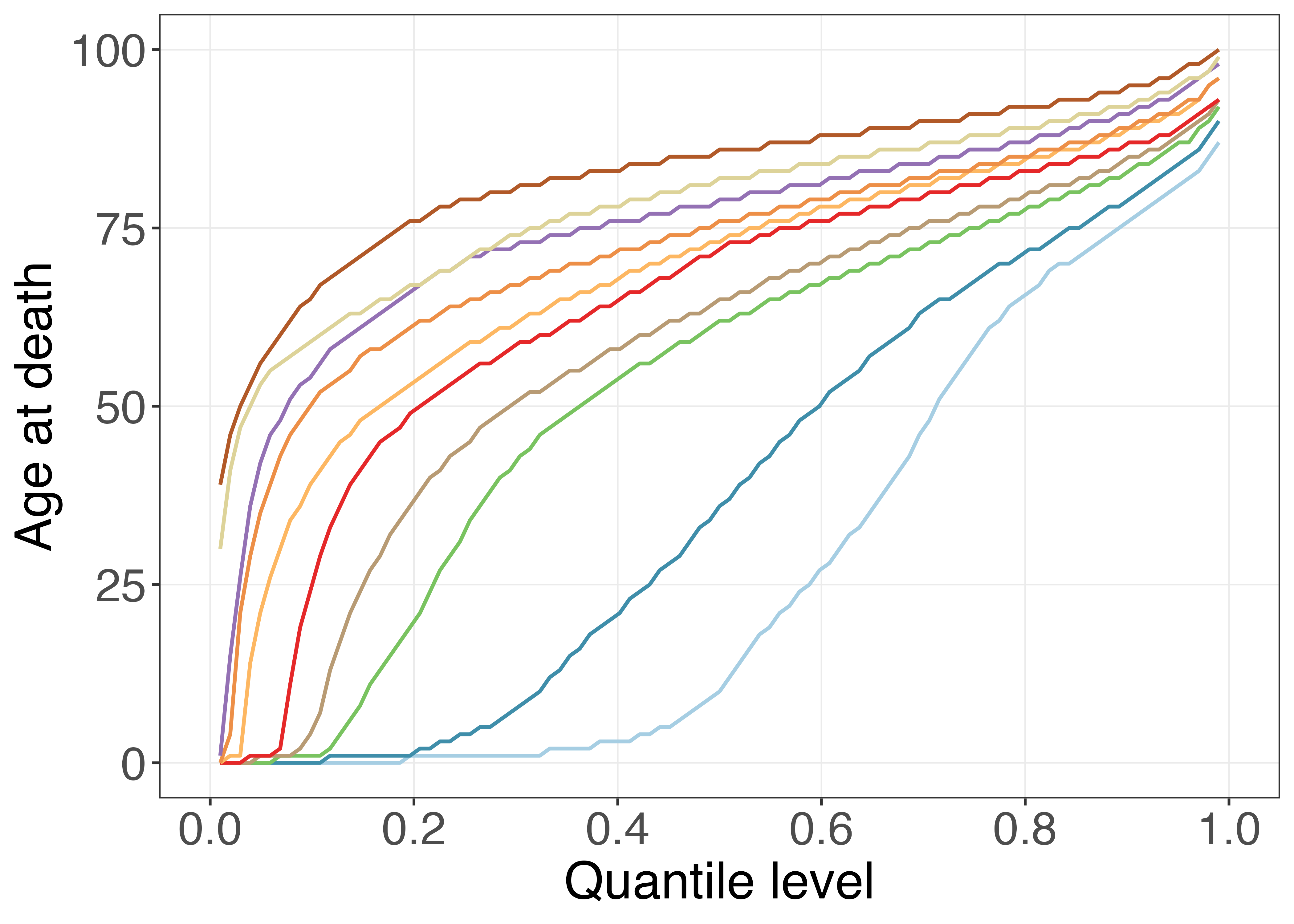}
    \end{minipage}\hspace{0.7cm}
    \begin{minipage}{0.51\textwidth}
        \centering
        \includegraphics[width=\linewidth]{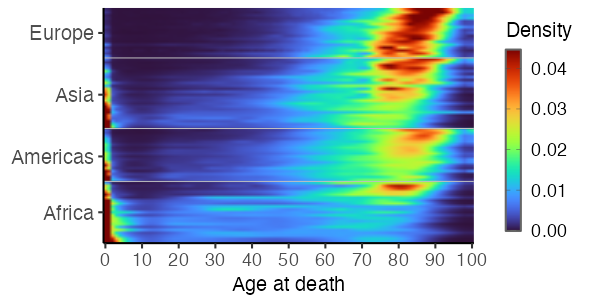}
    \end{minipage}
    \caption{Quantile functions (left) for the mortality distributions of 10 selected countries and a density heat map (right) for the mortality distributions of 80 countries.}
    \label{Figure: Mortality data}
\end{figure}

Figure~\ref{Figure: Mortality data} displays the mortality distributions from complementary perspectives. The left panel shows the quantile functions for 10 representative countries, while the right panel summarizes the full collection of densities for all 80 countries through a heat map. The plots reveal substantial heterogeneity not only in average longevity but also in infant mortality, the concentration of deaths at older ages, and the overall dispersion of the age-at-death distribution, supporting the use of distribution-valued responses in this application.

\begin{figure}[ht]
    \centering
    \includegraphics[width=0.48\textwidth]{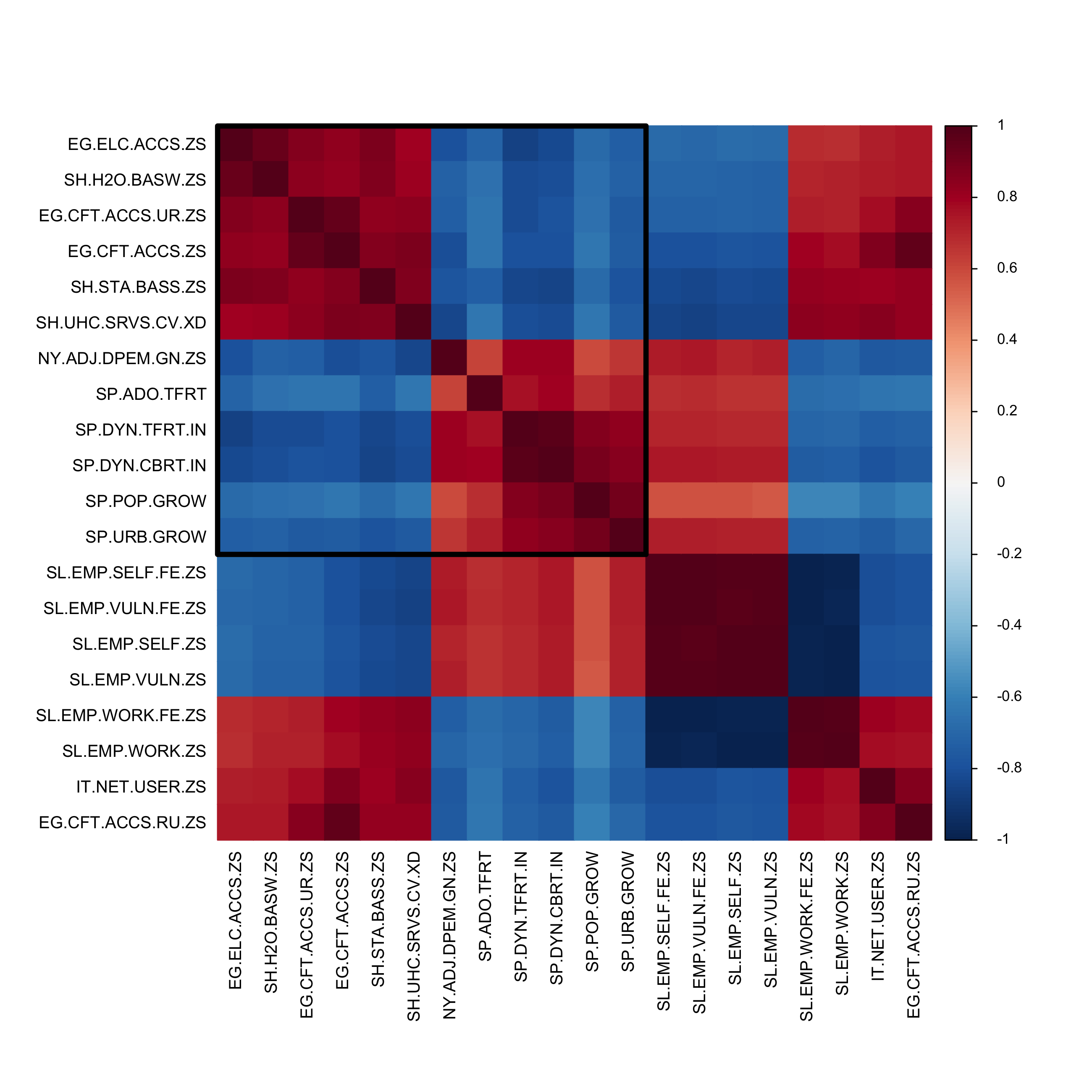}
    \includegraphics[width=0.48\textwidth]{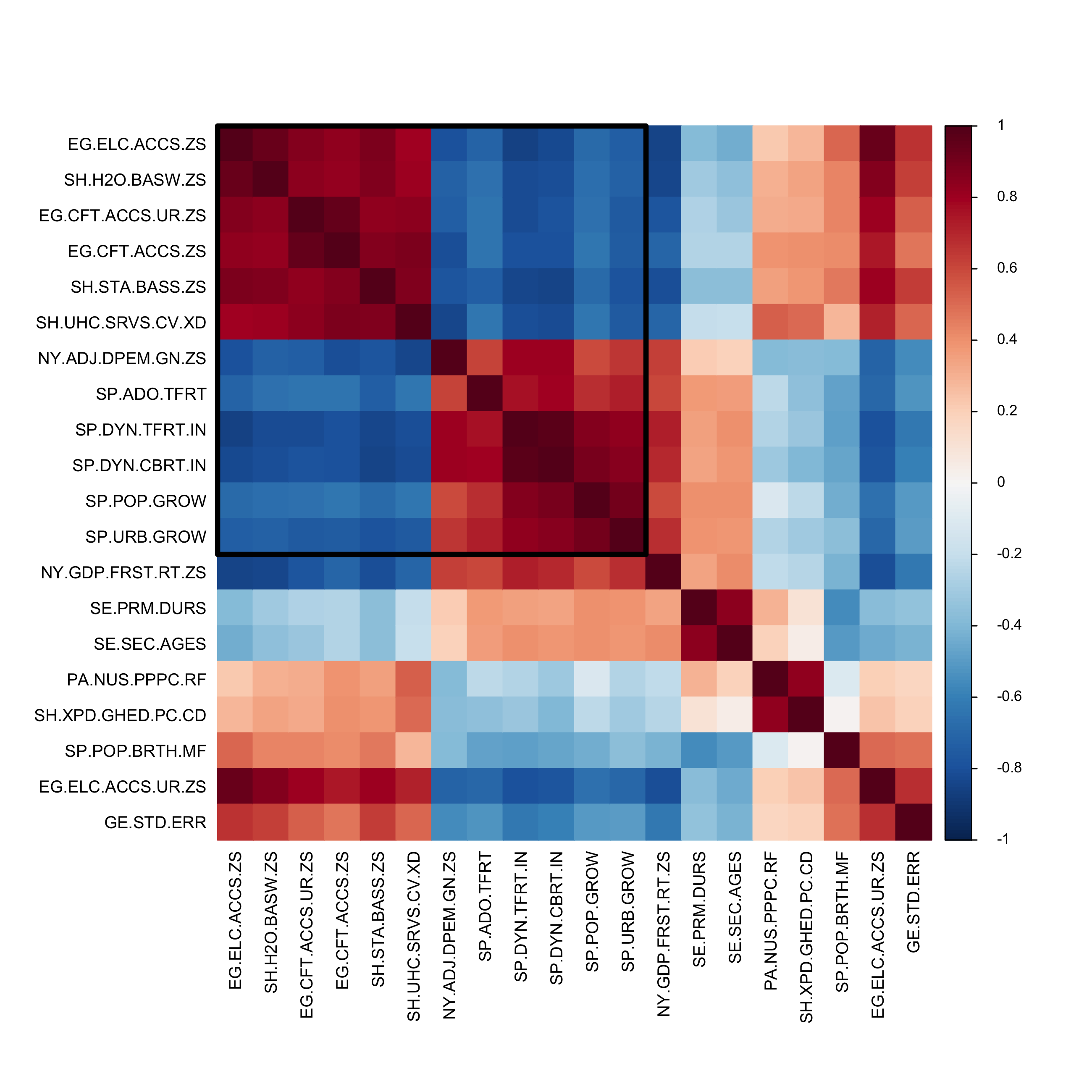}
    \caption{Correlation heatmaps of the top 20 indicators selected by Global Fr\'echet SIS (left) and Factor adjusted SIS (right).}
    \label{Figure: Top 20 indicators selected}
\end{figure}

Figure~\ref{Figure: Top 20 indicators selected} compares the correlation structure among the top 20 indicators selected by Global Fr\'echet SIS and by Factor adjusted SIS. Without factor adjustment, the screened set contains several near-duplicate employment measures and other closely related development proxies. Correspondingly, the left panel of Figure~\ref{Figure: Top 20 indicators selected} exhibits dense blocks of high correlation. This pattern suggests that ordinary marginal screening is largely tracking dominant common factors, such as overall socioeconomic development, rather than isolating predictors with distinct contributions to the mortality distribution. By contrast, the Factor adjusted SIS retains several core demographic and public-health indicators but also includes more differentiated variables such as forest rents and primary education duration. The full ranked lists for both methods are reported in Tables~\ref{Table: Top 20 indicators no factor} and \ref{Table: Top 20 indicators with factor adjustment}. The selected indicators suggest that female mortality profiles are associated with a combination of demographic behavior, access to basic services, environmental conditions, education, and public spending, rather than with a single generic development index.

\begin{table}[!t]
\centering
\footnotesize
\setlength{\tabcolsep}{4pt}
\renewcommand{\arraystretch}{0.65}

\caption{Top 20 indicators selected by Global Fr\'echet SIS without factor adjustment for the mortality analysis.}
\label{Table: Top 20 indicators no factor}
\vspace{-0.4em}

\begin{tabular}{@{}r l P{10.2cm} r@{}}
\toprule
Rank & Indicator & Indicator name & $\widehat R^2_{\oplus}$ \\
\midrule
1  & SP.DYN.CBRT.IN$^{\dagger}$     & Birth rate, crude (per 1,000 people) & 0.814 \\
2  & SP.DYN.TFRT.IN$^{\dagger}$     & Fertility rate, total (births per woman) & 0.775 \\
3  & SH.STA.BASS.ZS$^{\dagger}$     & People using at least basic sanitation services (\% of population) & 0.767 \\
4  & SH.UHC.SRVS.CV.XD$^{\dagger}$  & UHC service coverage index & 0.720 \\
5  & EG.CFT.ACCS.ZS$^{\dagger}$     & Access to clean fuels and technologies for cooking (\% of population) & 0.678 \\
6  & SP.URB.GROW$^{\dagger}$        & Urban population growth (annual \%) & 0.676 \\
7  & EG.ELC.ACCS.ZS$^{\dagger}$     & Access to electricity (\% of population) & 0.671 \\
8  & NY.ADJ.DPEM.GN.ZS$^{\dagger}$  & Adjusted savings, particulate emission damage (\% of GNI) & 0.658 \\
9  & IT.NET.USER.ZS                 & Individuals using the Internet (\% of population) & 0.657 \\
10 & EG.CFT.ACCS.UR.ZS$^{\dagger}$  & Access to clean fuels and technologies for cooking, urban (\% of urban population) & 0.633 \\
11 & EG.CFT.ACCS.RU.ZS              & Access to clean fuels and technologies for cooking, rural (\% of rural population) & 0.631 \\
12 & SP.POP.GROW$^{\dagger}$        & Population growth (annual \%) & 0.627 \\
13 & SH.H2O.BASW.ZS$^{\dagger}$     & People using at least basic drinking water services (\% of population) & 0.617 \\
14 & SL.EMP.SELF.FE.ZS              & Self-employed, female (\% of female employment) (modeled ILO estimate) & 0.600 \\
15 & SL.EMP.WORK.FE.ZS              & Wage and salaried workers, female (\% of female employment) (modeled ILO estimate) & 0.600 \\
16 & SL.EMP.VULN.FE.ZS              & Vulnerable employment, female (\% of female employment) (modeled ILO estimate) & 0.599 \\
17 & SP.ADO.TFRT$^{\dagger}$        & Adolescent fertility rate (births per 1,000 women ages 15--19) & 0.582 \\
18 & SL.EMP.WORK.ZS                 & Wage and salaried workers, total (\% of total employment) (modeled ILO estimate) & 0.578 \\
19 & SL.EMP.SELF.ZS                 & Self-employed, total (\% of total employment) (modeled ILO estimate) & 0.578 \\
20 & SL.EMP.VULN.ZS                 & Vulnerable employment, total (\% of total employment) (modeled ILO estimate) & 0.578 \\
\bottomrule
\end{tabular}

\vspace{0.3em}
\parbox{0.98\textwidth}{\footnotesize\emph{Notes:} Indicators marked with $^{\dagger}$ are also selected by the Factor adjusted SIS.}
\end{table}

\begin{table}[!t]
\centering
\footnotesize
\setlength{\tabcolsep}{3.5pt}
\renewcommand{\arraystretch}{0.72}

\caption{Top 20 indicators selected by Factor adjusted SIS for the mortality data.}
\label{Table: Top 20 indicators with factor adjustment}
\vspace{-0.4em}

\begin{tabular}{@{}r l P{7.3cm} r r@{}}
\toprule
Rank & Indicator & Indicator name & $\widehat \Delta_k$ &  $\widehat R_{\oplus,k}^2$ \\
\midrule
1  & SP.DYN.TFRT.IN$^{\dagger}$      & Fertility rate, total (births per woman) & 0.164 & 0.853 \\
2  & SP.DYN.CBRT.IN$^{\dagger}$      & Birth rate, crude (per 1,000 people) & 0.159 & 0.848 \\
3  & SH.STA.BASS.ZS$^{\dagger}$      & People using at least basic sanitation services (\% of population) & 0.135 & 0.824 \\
4  & EG.ELC.ACCS.ZS$^{\dagger}$      & Access to electricity (\% of population) & 0.134 & 0.823 \\
5  & EG.ELC.ACCS.UR.ZS               & Access to electricity, urban (\% of urban population) & 0.121 & 0.809 \\
6  & SP.POP.GROW$^{\dagger}$         & Population growth (annual \%) & 0.113 & 0.802 \\
7  & EG.CFT.ACCS.UR.ZS$^{\dagger}$   & Access to clean fuels and technologies for cooking, urban (\% of urban population) & 0.096 & 0.785 \\
8  & SP.URB.GROW$^{\dagger}$         & Urban population growth (annual \%) & 0.091 & 0.780 \\
9  & SP.POP.BRTH.MF                  & Sex ratio at birth (male births per female births) & 0.091 & 0.779 \\
10 & NY.ADJ.DPEM.GN.ZS$^{\dagger}$   & Adjusted savings, particulate emission damage (\% of GNI) & 0.088 & 0.776 \\
11 & NY.GDP.FRST.RT.ZS               & Forest rents (\% of GDP) & 0.086 & 0.775 \\
12 & SH.H2O.BASW.ZS$^{\dagger}$      & People using at least basic drinking water services (\% of population) & 0.080 & 0.769 \\
13 & EG.CFT.ACCS.ZS$^{\dagger}$      & Access to clean fuels and technologies for cooking (\% of population) & 0.079 & 0.768 \\
14 & SH.UHC.SRVS.CV.XD$^{\dagger}$   & UHC service coverage index & 0.078 & 0.767 \\
15 & SP.ADO.TFRT$^{\dagger}$         & Adolescent fertility rate (births per 1,000 women ages 15--19) & 0.077 & 0.766 \\
16 & SE.PRM.DURS                     & Primary education, duration (years) & 0.071 & 0.760 \\
17 & GE.STD.ERR                      & Government Effectiveness: Standard Error & 0.066 & 0.754 \\
18 & SE.SEC.AGES                     & Lower secondary school starting age (years) & 0.063 & 0.752 \\
19 & PA.NUS.PPPC.RF                  & Price level ratio of PPP conversion factor (GDP) to market exchange rate & 0.059 & 0.747 \\
20 & SH.XPD.GHED.PC.CD               & Domestic general government health expenditure per capita (current US\$) & 0.056 & 0.744 \\
\bottomrule
\end{tabular}

\vspace{0.3em}
\parbox{0.98\textwidth}{\footnotesize\emph{Notes:} Indicators marked with $^{\dagger}$ are also selected by the Global Fr\'echet SIS without factor adjustment.}
\end{table}

We also evaluate the screened variables as inputs for downstream multivariate Fr\'echet regression. Using forward stepwise selection with leave-one-out cross-validation to maximize adjusted Fr\'echet $R_{\oplus}^2$, the candidate set from Global Fr\'echet SIS yields a 5-variable model with adjusted $R^2_{\oplus}=0.912$ and $\mathrm{AIC}=266$. The candidate set from Factor adjusted SIS yields an 8-variable model with adjusted $R^2_{\oplus}=0.927$, and $\mathrm{AIC}=261$, leading to a better downstream fit. Together, these results show that factor adjustment does more than reduce redundancy at the screening stage. It also yields a stronger and more interpretable multivariate model.

\section{Concluding remarks}
\label{Section: Concluding remark}
In this paper, we proposed a Factor adjusted Fr\'{e}chet sure independence screening procedure for ultrahigh-dimensional predictors and metric-space-valued responses. The key idea is to combine latent factor recovery for the covariates with model comparison based on incremental Fr\'{e}chet $R_{\oplus}^2$, thus isolating the idiosyncratic contribution of each feature beyond the common factor structure. By working entirely within the Fr\'{e}chet regression framework, the proposed method avoids the response-residualization step that underlies Euclidean factor-adjusted procedures and is therefore applicable to a broad class of non-Euclidean responses. On the theoretical side, we established concentration for the sample Fr\'{e}chet coefficient of determination and uniform approximation of the feasible screening utilities to their population counterparts, leading to the sure screening and sure ranking properties. On the empirical side, the simulation studies showed clear gains over existing screening rules when predictors are strongly correlated, and the ADNI and mortality applications illustrated that factor adjustment can reduce redundant discoveries and uncover scientifically meaningful signals.

Possible extensions include data-adaptive thresholding, post-screening inference, and relaxing assumptions such as the boundedness of the metric space. It is also of interest to combine factor adjustment with iterative or conditional screening and with downstream sparse modeling, especially when signals are weak or jointly informative. We hope that the present work enriches the toolbox for high-dimensional analysis of random object data and provides a useful framework for multicollinearity-aware learning with non-Euclidean responses.


\bigskip
\begin{center}
{\large\bf SUPPLEMENTARY MATERIAL}
\end{center}

\textbf{Supplement to “Multicollinearity-agnostic feature screening for
non-Euclidean responses, a factor model approach”}: The online supplement includes the detailed toy exapmle, additional simulations, ancillary lemmas, and proofs of the main asymptotic results.

\bibliographystyle{apalike}
\bibliography{ref}

\end{document}